 \documentclass[preprint2,longabstract]{aastex}
\usepackage{graphicx}
\usepackage{epstopdf}

\newcommand{\fermi}{{\it Fermi} }
\newcommand{\swift}{{\it Swift} }

\slugcomment{To be submitted to ApJ}

\shorttitle{Observations of Gamma-Ray Sources by GBM}
\shortauthors{Case et al.}

\begin{document}

\title{First Results from \fermi GBM Earth Occultation Monitoring: Observations of Soft Gamma-Ray Sources Above 100 keV}


\author{G. L. Case, M. L. Cherry, J. C. Rodi}
\affil{Department of Physics and Astronomy, Louisiana State University,
    Baton Rouge, LA 70803, USA}
\email{case@phunds.phys.lsu.edu}

\author{P. Jenke\altaffilmark{1}, C. A. Wilson-Hodge}
\affil{Marshall Space Flight Center, Huntsville, AL 35182, USA}

\author{M. H. Finger, C. A. Meegan}
\affil{Universities Space Research Association, Huntsville, AL 35805, USA}

\author{A. Camero-Arranz}
\affil{National Space Science and Technology Center, Huntsville, AL 35805, USA}

\author{E. Beklen}
\affil{Physics Department, Middle East Technical University, 06531 Ankara, Turkey}
\affil{Physics Department, S\"uleyman Demirel University, 32260 Isparta, Turkey}

\author{P. N. Bhat, M. S. Briggs, V. Chaplin, V. Connaughton, W. S. Paciesas, R. Preece}
\affil{University of Alabama in Huntsville, Huntsville, AL 35899, USA}

\author{R. M. Kippen}
\affil{Los Alamos National Laboratory, Los Alamos, NM 87545}

\and

\author{A. von Kienlin, J. Greiner}
\affil{Max-Planck Institut f\"ur Extraterrestische Physik, 85748 Garching, Germany}

\altaffiltext{1}{NASA Postdoctoral Program Fellow}

\begin{abstract}
The NaI and BGO detectors on the Gamma-ray Burst Monitor (GBM) on \fermi are now being used for long-term monitoring of the hard X-ray/low energy gamma-ray sky. Using the Earth occultation technique as demonstrated previously by the BATSE instrument on the {\it Compton Gamma-Ray Observatory}, GBM can be used to produce multiband light curves and spectra for known sources and transient outbursts in the 8 keV to 1 MeV energy range with its NaI detectors and up to 40 MeV with its BGO detectors. Over 85\% of the sky is viewed every orbit, and the precession of the \fermi orbit allows the entire sky to be viewed every $\sim26$ days with sensitivity exceeding that of BATSE at energies below $\sim25$ keV and above $\sim1.5$ MeV. We briefly describe the technique and present preliminary results using the NaI detectors after the first two years of observations at energies above 100 keV.  Eight sources are detected with a significance greater than $7\sigma$: the Crab, Cyg X-1, SWIFT J1753.5-0127, 1E 1740-29, Cen A, GRS 1915+105, and the transient sources XTE J1752-223 and GX 339-4.  Two of the sources, the Crab and Cyg X-1, have also been detected above 300 keV.  
\end{abstract}


\section{Introduction}

The ability to monitor the gamma-ray sky continuously is extremely important.  The majority of gamma-ray sources are variable, exhibiting flares and transient outbursts on time scales from seconds to years. While there are currently several all-sky monitors in the hard X-ray energy range providing daily light curves, e.g. the All-Sky Monitor (ASM) on the {\it Rossi X-ray Timing Explorer} ({\it RXTE}) from 2--10 keV \citep{Levine1996}, the Gas Slit Camera (GSC) on the {\it Monitor of All-sky X-ray Image} ({\it MAXI}) from 1.5--20 keV \citep{Matsuoka2009}, and the Burst Alert Telescope (BAT) on \swift from 15--50 keV \citep{Gehrels2004}, there has not been an all-sky monitor in the low energy gamma-ray region since the Burst and Transient Source Experiment (BATSE) instrument on the {\it Compton Gamma-Ray Observatory} ({\it CGRO}), which was sensitive from 20--1800 keV \citep{Fishman1989}.

The gamma-ray satellite \fermi was launched on 2008 June 11 and commenced science operations on 2008 August 12. \fermi contains two instruments:  the Large Area Telescope (LAT), sensitive to gamma rays from $\sim20$ MeV to $\sim300$ GeV \citep{Atwood2009}; and the Gamma-ray Burst Monitor (GBM), which is sensitive to X-rays and gamma rays from 8 keV to 40 MeV \citep{Meegan2009}.  With its wide field of view, GBM can be used to provide nearly continuous full-sky coverage in the hard X-ray/soft gamma-ray energy range. It is the only instrument currently in orbit that can perform all-sky monitoring above 100 keV (but below the 30 MeV threshold of the \fermi LAT) with useable sensitivity. The \swift/BAT sensitivity drops off rapidly above 100 keV, and its energy range effectively ends at 195 keV. {\it INTEGRAL}, which has a relatively narrow field of view, cannot make continuous observations of a large number of individual sources. Also, GBM is not limited by solar pointing constraints, as are most other instruments, which allows the monitoring of sources at times during which other instruments cannot. 

The Earth occultation technique, used very successfully with BATSE, has been adapted to GBM to obtain fluxes for an input catalog of known or potential sources. A catalog of 82 sources is currently being monitored and regularly updated. This catalog contains predominantly Galactic x-ray binaries, but also includes the Crab, the Sun, two magnetars, five active galactic nulcei (AGN), and two cataclysmic variables.  At energies above 100 keV, six persistent sources (the Crab, Cyg X-1, SWIFT J1753.5-0127, 1E 1740-29, Cen A, GRS 1915+105) and two transient sources (GX 339-4 and XTE J1752-223) have been detected in the first two years of observations. In section~\ref{sec:obs}, we briefly describe the GBM instrument and outline the Earth occultation technique as applied to GBM, in section~\ref{sec:res} we present the light curves for the eight sources, and in section~\ref{sec:dis} we discuss the results, the GBM capabilities and future work. 

\section{GBM and the Earth Occultation Technique \label{sec:obs}}

GBM consists of 14 detectors: 12 NaI detectors, each 12.7 cm in diameter and 1.27 cm thick; and two BGO detectors, 12.7 cm in diameter and 12.7 cm thick.  The NaI detectors are located on the corners of the spacecraft, with six detectors oriented such that the normals to their faces are perpendicular to the z-axis of the spacecraft (the LAT is pointed in the $+z$-direction), four detectors pointed at $45^{\circ}$ from the z-axis, and 2 detectors pointed $20^{\circ}$ off the z-axis.  Together, these 12 detectors provide nearly uniform coverage of the unocculted sky in the energy range from 8 keV to 1 MeV.  The two BGO detectors are located on opposite sides of the spacecraft and view a large part of the sky in the energy range $\sim150$ keV to $\sim40$ MeV. It should be noted that none of the GBM detectors have direct imaging capability.

\begin{figure}[t]
\includegraphics[width=77mm]{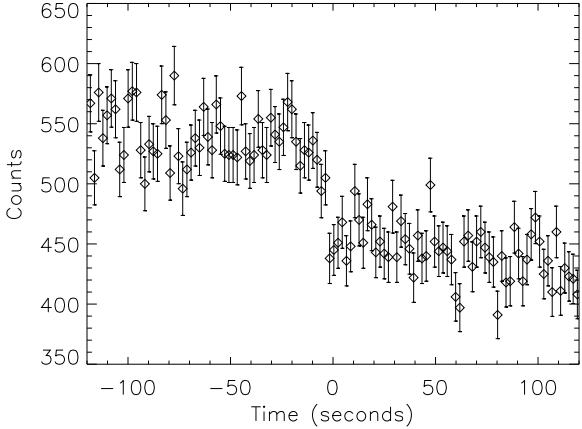}
\caption{\label{crabstep}Single Crab occultation step seen in the CTIME raw counts data of a single GBM NaI detector (NaI 2) in the 12--25 keV band with 2.048-second time bins. The Crab was $4.5^{\circ}$ from the normal to the detector. The time window is centered on the calculated occultation time for 100 keV.}
\end{figure}

Known sources of gamma-ray emission can be monitored with non-imaging detectors using the Earth occultation technique, as was successfully demonstrated with BATSE \citep{Ling2000,Harmon2002,Harmon2004}. When a source of gamma rays is occulted by the Earth, the count rate measured by a detector will drop, producing a step-like feature. When the source reappears from behind the Earth's limb, the count rate will increase, producing another step. The occultation has a finite transition time due to the effect of absorption in the Earth's atmosphere. The shapes of the individual occultation steps depend on energy and occultation angle. The occultation angle, $\beta$, is defined as the elevation angle of the source being occulted with respect to the plane of the \fermi orbit. The transmission through the atmosphere as a function of time is modeled as $T(t) = \exp[-\mu(E) A(h)]$, where $\mu(E)$ is the mass attenuation coefficient of gamma rays at energy $E$ in air and $A(h)$ is the air mass along the line of sight at a given altitude $h(t)$ based on the \citet{atm1976}. This requires instantaneous knowledge of the spacecraft position, the direction to the source of interest as seen from the spacecraft, and a model of the Earth that includes its oblateness. \fermi was launched into a $i = 25.6^\circ$ inclination orbit at an altitude of 555 km. The orbital period is 96 minutes, and individual occultation steps last for $\sim8/\cos\beta$ seconds.  Figure~\ref{crabstep} shows a single step due to a Crab occultation in the count rate in the 12--25 keV band of a single GBM NaI detector observing the occultation nearly face-on.

The diameter of the Earth as seen from \fermi is $\approx 135^\circ$, so roughly 30\% of the sky is occulted by the Earth at any one time.  One complete orbit of the spacecraft allows over 85\% of the sky to be observed.  The precession of the orbital plane allows the entire sky to be occulted every $\sim26$ days (half the precession period for the \fermi orbit), though the exposure is not uniform.

The Earth occultation technique was developed for BATSE using two separate approaches, one at the NASA Marshall Space Flight Center \citep{Harmon2002} and the other at the NASA Jet Propulsion Laboratory \citep{Ling2000}. For GBM, we follow the \citet{Harmon2002} approach. 

The primary difference in the implementation of the occultation technique between GBM and BATSE arises from the different pointing schemes of the respective missions. {\it CGRO} was three-axis stabilized for each viewing period, which typically lasted for two weeks. This meant that a source remained at a fixed orientation with respect to the detectors through an entire viewing period. In contrast, \fermi scans the sky by pointing in a direction $35^\circ$ (August 2008--September 2009) or $50^\circ$ (October 2009--present) north of the zenith for one orbit; it then rocks to $35^\circ$ or $50^\circ$ south for the next orbit, continuing to alternate every orbit unless the spacecraft goes into a pointed mode (which occurs rarely). In addition, the spacecraft performs a roll about the z-axis as it orbits.  Because the orientation of a source with respect to the GBM detectors varies as a function of time, the detector response as a function of angle must be accounted for. A detailed instrument modeling and measurement program has been used to develop the GBM  instrument response as a function of direction \citep{Hoover2008,Bissaldi2009}, which is incorporated into the occultation analysis. It should be noted that the GBM occultation sensitivity exceeds that of BATSE at energies below $\sim25$ keV and above $\sim1.5$ MeV, since GBM has only a Be window on the NaI detectors instead of the plastic scintillator, aluminum honeycomb, and aluminum window that covered the front of the BATSE scintillators \citep{Case2007}.  Due to its larger area, BATSE was more sensitive than GBM between 25 keV and 1.5 MeV.

GBM has two continuous data types: CTIME data with nominal 0.256-second time resolution and 8-channel spectral resolution and CSPEC data with nominal 4.096-second time resolution and 128-channel spectral resolution. The results presented in this paper use the low-spectral resolution CTIME data. Detailed spectral analyses using the higher resolution CSPEC data are reserved for future work.
The data are selected to remove gamma-ray bursts, solar flares, electron precipitation events, cosmic ray events, terrestrial gamma-ray flashes, etc. The occultation technique requires an input 
catalog of predetermined source locations, and currently we are monitoring 82 sources. For each day, the occultation times for each source are calculated using the known spacecraft positions. The time of the occultation step is taken to be the time for which the transmission of a 100 keV gamma ray through the atmospheric column is 50\%.  The time at which the atmospheric transmission reaches 50\% is energy dependent, e.g. for energies less than 100 keV, a setting step will occur earlier (see Fig.~\ref{crabstep}). This energy dependence is accounted for in the calculation of the atmospheric transmission function, $T(t)$. For each occultation step, a 4-minute window is defined that is centered on the 100 keV occultation time.  For each energy band, the count rates in the window are fit separately for each detector viewing the source of interest. In each of these detectors, the count rates are fitted with a quadratic background plus source models for the source of interest and each interfering source that appears in the window. The source models consist of $T(t)$ and a time-dependent model count rate, derived from the time-dependent detector response convolved with an assumed source spectrum.  Each source model is multiplied by a scaling factor, and the source flux is then computed by a joint fit to the scaling factors across all detectors in the fit. The best-fit scaling factor is then multiplied by the assumed source flux model integrated over the energy band to obtain the photon flux.

Up to 31 occultation steps are possible for a given source in a day, and these steps are summed to get a single daily average flux. This technique can be used with either the NaI or BGO detectors, though the analysis presented here uses only the NaI detectors. A more complete description of the GBM implementation of the occultation technique will be given in \citet{Wilson2010}.

\section{Results \label{sec:res}}

\begin{deluxetable}{lccccccccc}
\tabletypesize{\footnotesize}
\tablecaption{Fluxes and Significances in GBM Broad High Energy Bands\label{Flux_table}}
\tablehead{
\colhead{} & \multicolumn{3}{c}{50--100 keV} & \multicolumn{3}{c}{100--300 keV} & \multicolumn{3}{c}{300--500 keV} \\
& Flux & Error & Signif. & Flux & Error & Signif. & Flux & Error & Signif. \\
& (mCrab) & (mCrab) & ($\sigma$) & (mCrab) & (mCrab) & ($\sigma$)& (mCrab) & (mCrab) & ($\sigma$) 
}
\startdata
Cyg X-1           & 1151.0 &  3.7 & 312 & 1130.7 &  6.9 & 163 &  529.0 &  49.5 & 10.7 \\
Crab              & 1000.0 &  3.3 & 307 & 1000.0 &  6.3 & 158 & 1000.0 &  48.0 & 20.9 \\
XTE J1752-223\tablenotemark{a}  & 730.8  & 14.2 &  51 &  563.1 & 26.7 &  21 &  226.1 & 204.4 &  1.1 \\
Cen A             &  72.4  &  3.6 &  20 &  104.2 &  6.7 &  16 &   $<96.2$\tablenotemark{c} & \ldots & \ldots \\
SWIFT 1753.5-0127 & 121.0  &  4.4 &  28 &  126.8 &  8.2 &  15 &  104.5 &  62.0 &  1.7 \\
1E 1740-29        & 116.3  &  4.7 &  25 &  92.3  &  8.8 &  11 &  126.8 &  65.0 &  2.0 \\
GRS 1915+105      & 128.1  &  3.6 &  35 &  54.9  &  6.8 &  8.0 &   $ <101.2$\tablenotemark{c} & \ldots & \ldots \\
GX 339-4\tablenotemark{b}    & 399.4  & 18.3 &  22 &  249.5 & 33.8 &   7.4 & $<507.4$\tablenotemark{c} & \ldots & \ldots \\
\enddata
\tablenotetext{a}{Fluxes are given for MJD 55129-55218 when XTE J1752-223 was flaring.}
\tablenotetext{b}{Fluxes are given for MJD 55244-55289 when GX 339-4 was flaring.}
\tablenotetext{c}{$2\sigma$ upper limit.}
\end{deluxetable}

In \citet{Wilson2009a}, the measured GBM 12--50 keV light curves are compared to the \swift BAT 15--50 keV light curves for several sources over the same time intervals, and it is seen that the fluxes measured by the two instruments compare well. At energies above the $\sim195$ keV upper energy limit of the \swift 22-month catalog \citep{Tueller2010}, however, the GBM observations provide the only wide-field monitor available for the low energy gamma-ray sky. Of the catalog sources being monitored with GBM, six persistent sources have been detected above 100 keV with a statistical significance of at least $7\sigma$ after two years of observations, as well as two transient sources.

Table \ref{Flux_table} gives the fluxes averaged over all 730 days from 2008 August 12 (MJD 54690, the beginning of science operations) to 2010 August 11 (MJD 55419) for the persistent sources, and over all of the days of the flares for the transient sources. Also given are the significances for each energy band. The errors are statistical only.  The sources are sorted by their detection significance in the 100--300 keV band.

\subsection{Persistent Sources}
The six persistent sources Crab, Cyg X-1,  Cen A, GRS 1915+105, 1E 1740-29, and Swift J1753.5-0127 are detected by GBM at energies above 100 keV.  In Figures~\ref{Crab}--\ref{Swift} we show light curves for these sources generated from the GBM data in several broad energy bands with five-day resolution. These persistent sources demonstrate the capabilities of the GBM Earth occultation monitoring.

\subsubsection{Crab}
\begin{figure}[t]
\includegraphics[width=78mm]{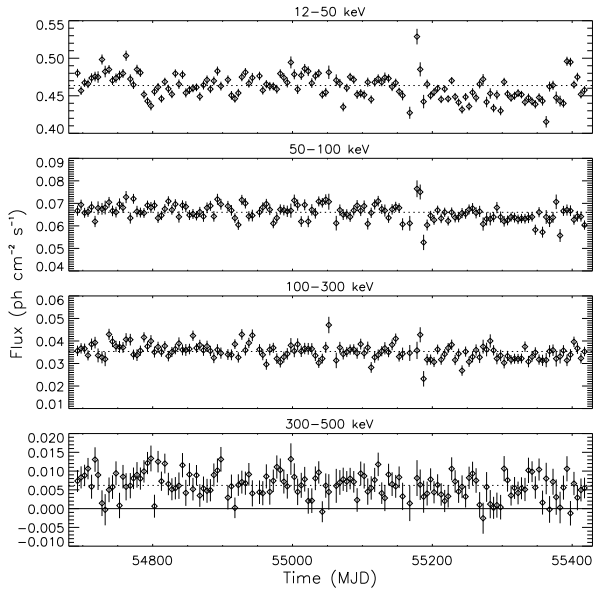}%
\caption{\label{Crab}GBM light curve for the Crab. The horizontal scale is in modified Julian days over the 730 day GBM exposure period, and has been binned 5 days per data point. The dashed horizontal lines show the average flux in each of four energy bands increasing from top to bottom.  In the bottom plot, the solid line marks the zero flux level. Note that the apparent ``flare'' near MJD 55180 is due to a giant outburst in the nearby accreting pulsar A0535+262.}
\end{figure}

The Crab emission in the hard X-ray/low energy gamma-ray regime contains a combination of pulsar and pulsar wind nebula contributions.  Figure~\ref{Crab} shows the light curves measured by GBM in four broad energy bands from 12 keV up to 500 keV. 
The spectrum in this regime has been shown by analysis of BATSE occultation data \citep{Much1996,Ling2003b} and data from SPI on board {\it INTEGRAL} \citep{Jourdain2009} to agree with the spectrum measured with other instruments at lower X-ray energies, and then to steepen near 100 keV. Results of the BATSE analysis can be described by a broken power law, while results of the SPI analysis suggest a smoothly steepening spectrum. The BATSE analysis further noted a distinct hardening of the spectrum near 650 keV, although this has not been confirmed by {\it INTEGRAL} or the COMPTEL instrument on {\it CGRO} \citep{Kuiper2001}. 

The {\it INTEGRAL} spectral measurements are consistent with either a smoothly steepening spectrum or a spectrum of the form $ F = 6.6 \times 10^{-4} (E/100 {\rm ~keV})^{-\alpha}$ photons cm$^{-2}$ s$^{-1}$ keV$^{-1}$, where $\alpha = 2.07 \pm 0.01$ for $E < 100$ keV and $\alpha = 2.23 \pm 0.02$ for $E > 100$ keV \citep{Jourdain2009}. This corresponds to a (50--100 keV)/(12--50 keV) flux ratio of $R_{50} = 0.145$ for {\it INTEGRAL} compared to $0.142 \pm 0.001$ for GBM. The (100--300 keV)/(12--50 keV) flux ratio $R_{100} = 0.041$ corresponding to the {\it INTEGRAL} spectrum compares to the GBM value of $0.076 \pm 0.001$, and the (300--500 keV)/(12--50 keV) flux ratio $R_{300} = 0.007$ for {\it INTEGRAL} corresponds to the GBM value of $0.013 \pm 0.001$. The GBM measurements suggest a somewhat flatter spectrum than that derived from {\it INTEGRAL}, particularly above 100 keV, and are best described by a spectrum with $\alpha \sim 2.09-2.12$.

These results demonstrate that GBM is able to see significant emission above 300 keV at a level consistent with the canonical hard spectrum of the Crab. Future analysis with the finer spectral resolution of the GBM CSPEC data will allow a better determination of the the break energy and spectral index above the break.

\subsubsection{Cyg X-1} 

\begin{figure}[t]
\includegraphics[width=78mm]{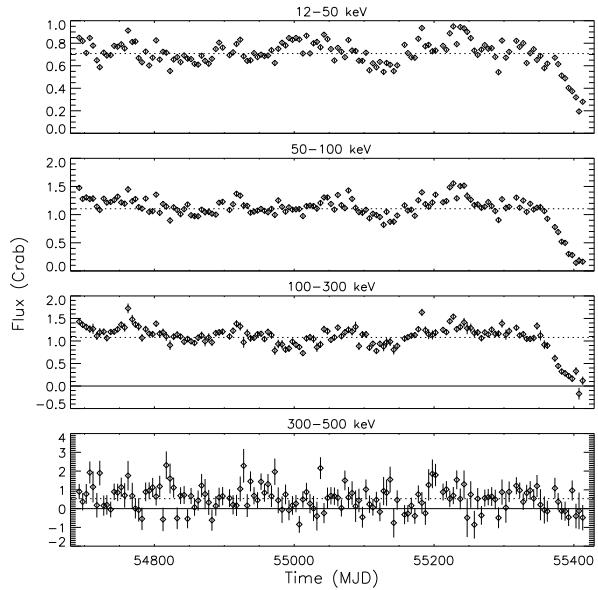}
\caption{\label{CygX1} The GBM light curve for Cyg X-1 over 730 days. The light curve has been binned 5 days per data point.  The fluxes are in Crab units, and the dashed and solid lines mark the average flux and zero flux levels, respectively.}
\end{figure}

Cygnus X-1 is a high-mass X-ray binary and was one of the first systems determined to contain a black hole \citep{Bolton1972,Paczynski1974}.  The X-ray emission is bimodal, with the $>10$ keV emission anticorrelated with the $<10$ keV emission \citep{Dolan1977}. It has been observed to emit significant emission above 100 keV including a power law tail extending out to greater than 1 MeV \citep{McConnell2000,Ling2005b}. Based on BATSE occultation analysis, \citet{Ling2005b} have shown that in the high gamma-ray intensity (hard) state, the spectrum consists of a Comptonized shape below 200--300 keV with a soft ($\Gamma > 3$) power-law tail extending to at least 1 MeV. In the low-intensity (soft) state, however, the spectrum takes on a different shape: In this case, the entire spectrum from 30 keV to 1 MeV is characterized by a single power law with a harder index $\Gamma \sim 2 - 2.7$. Similar behavior has been reported for two low-mass X-ray binaries (LMXBs) that also contain black holes, the transient gamma-ray sources GRO J0422+32 \citep{Ling2003a} and GRO J1719-24 \citep{Ling2005a}. 

Figure \ref{CygX1} shows the GBM light curves.  The light curves show significant variability with emission above 300 keV up until about MJD 55355.  Starting at about MJD 55355, the 100--300 keV band emission began to decrease \citep{WilsonCase2010}, dropping from an average level of about 1200 mCrab down to nearly undetectable levels on MJD 55405--55406.  On MJD 55374 MAXI detected a rapid rise in the soft 2--4 keV band \citep{Negoro2010b}, which combined with the decrease in the low energy gamma-ray flux indicated a transition to a thermally-dominated soft state. As of MJD 55419 the one-day average GBM light curves show that the 12--50 keV flux has begun to rise, while the 100--300 keV flux remains at a low level of $\approx 150$ mCrab, consistent with the $\gamma_0$ state of \citet{Ling1997}. We will continue to monitor Cyg X-1 during the thermally-dominated state and follow its transition back to the low/hard state.

The GBM light curves (Fig.~\ref{CygX1}) reveal significant emission above 300 keV, consistent with the power law tail observed when Cyg X-1 is in its low/hard state. 
The 50--100 keV flux level observed by GBM over the full observation period (Table~\ref{Flux_table}) is 1.15 Crab, consistent with the BATSE high gamma-ray state ($\gamma_2$ of \citet{Ling1987,Ling1997}). 
The observed GBM flux ratios are $R_{50} = 0.226 \pm 0.001$, $R_{100} = 0.119 \pm 0.001$, and $R_{300} = 0.010 \pm 0.001$, inconsistent with a single power law. A single power law with $\Gamma = 2$ would yield flux ratios 0.158, 0.105, and 0.021. Instead, the GBM ratios suggest a spectrum that appears significantly flatter at low energies and steeper at high energies, consistent with the behavior reported earlier from the BATSE analysis. 

\subsubsection{Cen A} 

\begin{figure}[t]
\includegraphics[width=78mm]{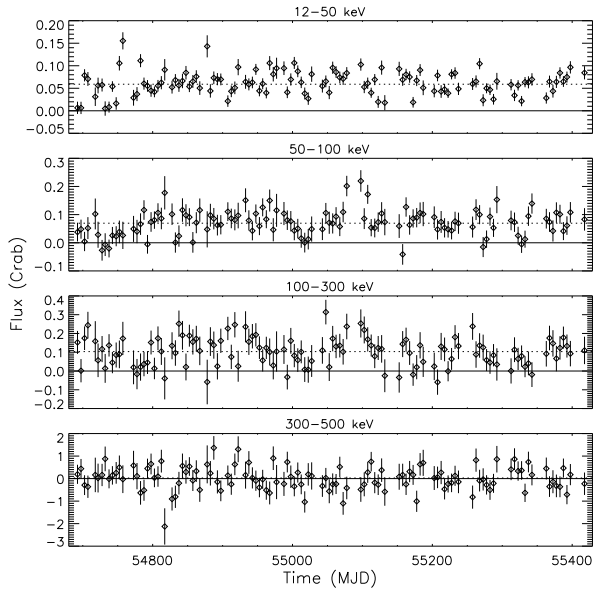}%
\caption{\label{CenA} The GBM light curve for Cen A. The light curve has been binned 5 days per data point.  The fluxes are in Crab units, and the dashed and solid lines mark the average flux and zero flux levels, respectively.}
\end{figure}

The relatively nearby radio galaxy Cen A is a Seyfert 2 galaxy that is the brightest AGN in hard X-rays/low energy gamma rays. It has powerful jets aligned at approximately $70^\circ$ from the line of sight and is seen to vary on time scales of tens of days to years. It has been observed at hard X-ray energies by {\it OSSE} \citep{Kinzer1995}, {\it INTEGRAL} and {\it RXTE} \citep{Rothschild2006}, and at energies $>1$ MeV by COMPTEL \citep{Steinle1998}. The observations below 150 keV are consistent with a hard spectrum with a power law index $\Gamma \sim 1.8-1.9$. The combined OSSE and COMPTEL data are consistent with a steepening of the spectrum at 150 keV to $\Gamma \sim 2.3$, with the spectrum then extending unbroken to beyond 10 MeV.  

The GBM light curve for Cen A is shown in Fig.~\ref{CenA}.  Because Cen A is relatively far below the equatorial plane, with a declination $\delta = -43^{\circ}$, its beta angle (which ranges between $\delta \pm i$) can be larger than the half-angle size of the Earth as seen from \fermi ($\beta_{\rm{earth}} \approx \pm 67^{\circ}$).  When this happens, Cen A is not occulted.  This causes periodic gaps in the light curve, with the period of the gaps equal to the precession period of the orbit.
 
The fluxes as measured by GBM and given in Table \ref{Flux_table} are consistent with the hard spectrum measured by previous instruments. The flux ratios measured by GBM are $R_{50} =0.168 \pm 0.008$ and $R_{100} = 0.134 \pm 0.008$ respectively. An unbroken $\Gamma = 1.9$ power law up to 300 keV would result in flux ratios of 0.178 and 0.129, respectively. A $\Gamma = 1.9$ power law extending up to 500 keV would result in a (300--500 keV)/(12--50 keV) flux ratio of 0.028, while GBM measures essentially no flux above 300 keV, $R_{300} = 0.006 \pm 0.010$, consistent with a steepening or cut off somewhere near 300 keV. The GBM results appear to be consistent with the steepening seen in the OSSE-COMPTEL spectra. 

\subsubsection{GRS 1915+105} 

\begin{figure}[t]
\includegraphics[width=78mm]{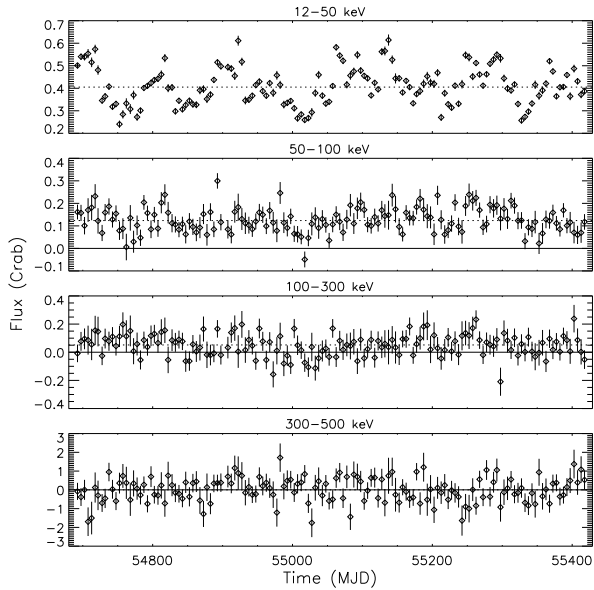}%
\caption{\label{GRS1915}GRS 1915+105 light curve. The light curve has been binned 5 days per data point.  The fluxes are in Crab units, The fluxes are in Crab units, and the dashed and solid lines mark the average flux and zero flux levels, respectively.}
\end{figure}

The galactic microquasar GRS 1915+105 is a LMXB with the compact object being a massive black hole \citep{Greiner2001}. It was highly variable over the 9-year observation period of the {\it CGRO} mission \citep{Paciesas1995,Case2005} with significant emission observed out to $\sim 1$ MeV \citep{Zdziarski2001,Case2005}. Combining the BATSE data from multiple outbursts yields a spectrum best fit by a broken power law, with  spectral index $\Gamma \sim 2.7$ below 300 keV  flattening to $\Gamma \sim 1.5$ above 300 keV. The spectrum derived from BATSE data shows no evidence of the thermal spectrum seen in Cyg X-1. By contrast, observations with {\it INTEGRAL}/SPI in the 20--500 keV energy range \citep{Droulans2009} showed evidence for a time-variable thermal Comptonization component below $\sim 100$ keV along with a relatively steady, hard power law at higher energies, indicating that different emission regions are likely responsible for the soft and hard emission. 

The GBM daily fluxes integrated over 730 days (Table \ref{Flux_table}) show significant emission above 100 keV, consistent with the relatively hard power law spectrum seen in BATSE and SPI data. The GBM light curve (Fig.~\ref{GRS1915}) shows distinct variability below 100 keV, with statistics above 100 keV insufficient to determine the level of variability of the emission.  
The flux ratios observed by GBM ($R_{50} = 0.044 \pm 0.001$ and $R_{100} = 0.010 \pm 0.001$) are close to the flux ratios expected from a power law spectrum with $\Gamma \sim 3$ ($R_{50} = 0.046$ and $R_{100} = 0.014$).

\subsubsection{1E 1740-29} 

\begin{figure}[t]
\includegraphics[width=78mm]{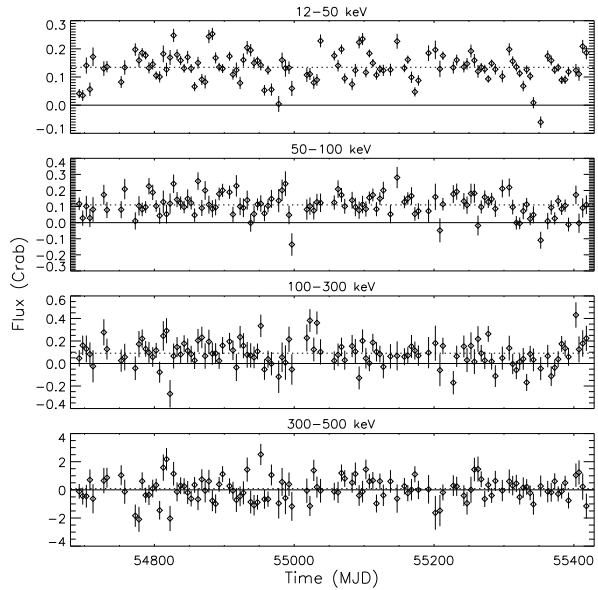}%
\caption{\label{1e1740}The GBM light curve for 1E 1740-29. The light curve has been binned 5 days per data point.  The fluxes are in Crab units, and the dashed and solid lines mark the average flux and zero flux levels, respectively.}
\end{figure}

The black hole candidate 1E 1740-29 (also known as 1E 1740.7-2942) is a LMXB very near the Galactic Center. With a large double-ended radio jet, it was the first source identified as a microquasar, and spends most of its time in the low/hard state \citep{Mirabel1992}.  {\it INTEGRAL} observations indicate the presence of significant emission up to at least 500 keV with a steepening of the spectrum near 140 keV \citep{Bouchet2009}. The spectrum can be modeled either with a thermalized Compton spectrum and a high energy power law tail, or with two superimposed thermal Compton components. Evidence for a broad 511 keV line observed by SIGMA \citep{Bouchet1991, Sunyaev1991} suggests that 1E 1740-29 may be a source of positrons.  

The GBM results (Fig.~\ref{1e1740}) are consistent with the high energy component observed when 1E 1740-29 is in the low/hard state. Below 100 keV and above 300 keV, GBM sees approximately $20-50\%$ higher flux than {\it INTEGRAL}, while in the 100--300 keV band, GBM observes approximately 90\% of the level reported by {\it INTEGRAL}. 


\subsubsection{SWIFT J1753.5-0127} 

\begin{figure}[t]
\includegraphics[width=78mm]{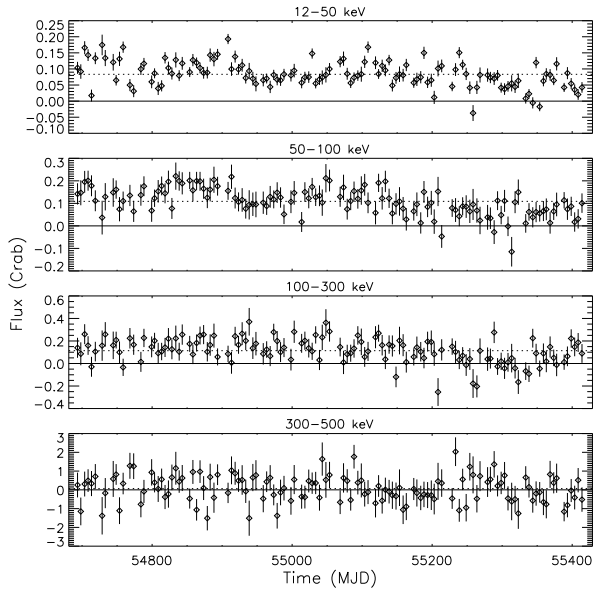}%
\caption{\label{Swift} The GBM light curve for SWIFT J1753.5-0127. The light curve has been binned 5 days per data point. The fluxes are in Crab units, with the average flux (dashed lines) and zero flux (solid lines) levels shown.}
\end{figure}

The X-ray nova SWIFT J1753.5-0127 (Fig.~\ref{Swift}) is a LMXB with the compact object likely being a black hole \citep{Miller2006,Bel2007}.  \swift discovered this source when it observed a large flare in 2005 July \citep{Palmer2005}.  The source did not return to quiescence but settled into a low intensity hard state \citep{Miller2006}. {\it INTEGRAL} observations \citep{Bel2007} showing emission up to $\sim 600$ keV were compatible with thermal Comptonization modified by reflection, with evidence for separate contributions from a jet, disk, and corona. BATSE occultation measurements from 1991--2000 showed no significant emission from this source above 25 keV \citep{Case2010}.  

The GBM results are consistent with this source still remaining in a hard state, with significant emission in excess of 100 mCrab above 100 keV. The light curves show that the emission from the higher energy bands declined beginning about MJD 55200, and that it increased again beginning about MJD 55325 and is currently at or just below its two-year average. The spectrum is inconsistent with a single power law, and future work using the GBM CSPEC data will allow a more detailed analysis of the spectrum. 
We will continue to monitor this source while it is in the low/hard state.

\subsection{Transient Sources}

\subsubsection{XTE J1752-223}

\begin{figure}[t]
\includegraphics[width=78mm]{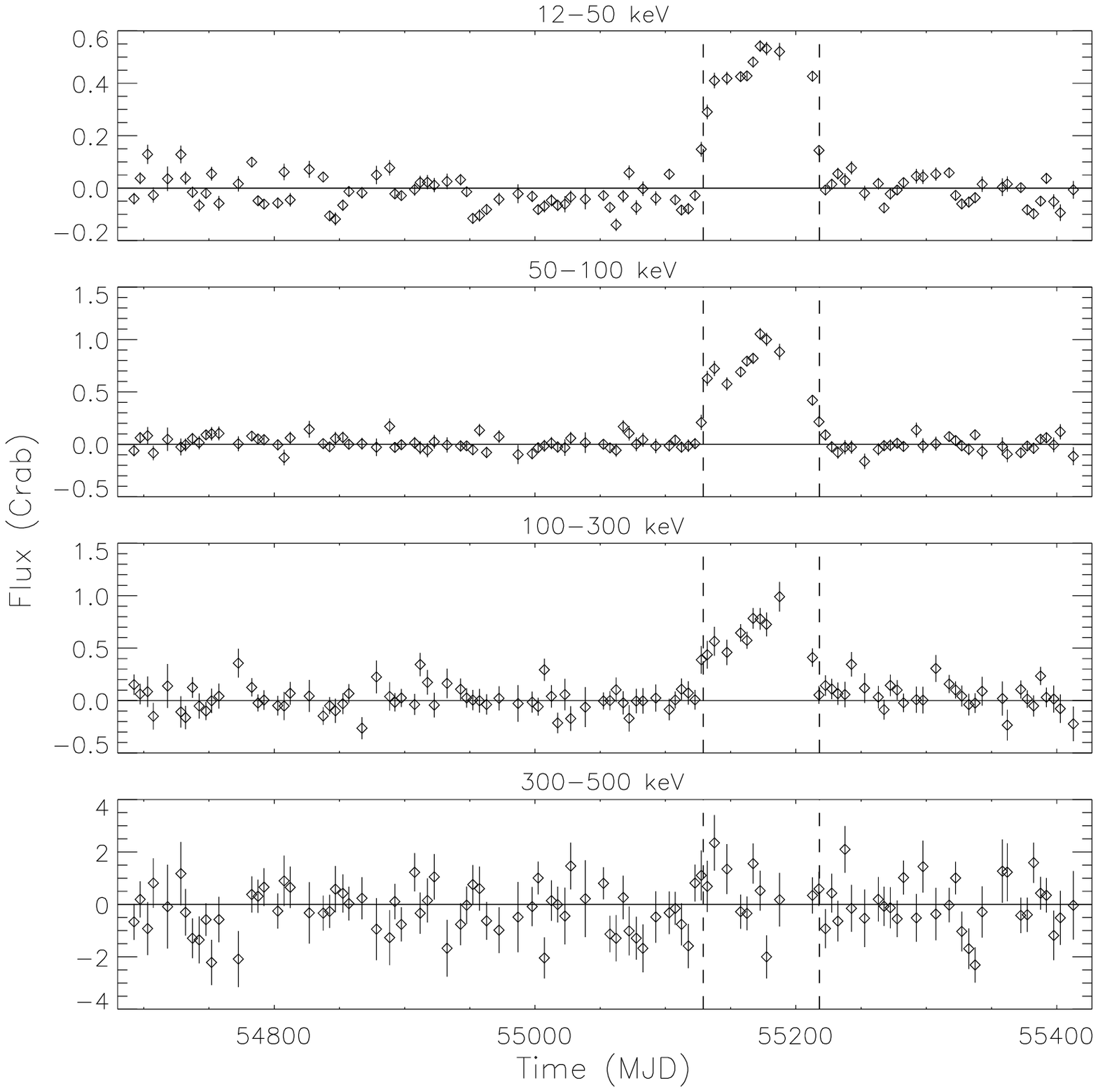}%
\caption{\label{XTEJ1752} The GBM light curve for XTE J1752-223. The light curve has been binned 5 days per data point.  The vertical dashed lines at MJD 55129 and MJD 55218 mark the flaring region used to derive the average fluxes in Table~\ref{Flux_table}. The fluxes are in Crab units, and the horizonial solid lines mark the zero flux levels.}
\end{figure}

The new transient black hole candidate XTE J1752-223, discovered by {\it RXTE} \citep{Markwardt2009b}, was observed by GBM to rise from undetectable on 2009 October 24 (MJD 55128) to $511\pm 50$ mCrab (12--25 keV), 
$570\pm 70$ mCrab (25--50 keV), $970\pm 100$ mCrab (50--100 keV), and $330\pm 100$ mCrab (100--300 keV) on 2009 November 2 \citep{Wilson2009a,Wilson2009b}. The light curve is 
variable, especially in the 12--25 keV band, where the flux initially rose to about 240 mCrab (25-28 Oct), suddenly dropped to non-detectable on 
October 29-30, then rose again during the period October 31 to November 2 (MJD 55135--55137). The flux remained relatively constant until November 25 (MJD 55160) when it began to rise again, peaking in the high energies on 2009 December 20 (MJD 55185).  After an initial slow decline, the high energy flux rapidly declined back to the pre-flare levels. The light curve for the entire mission to date, with 5-day resolution, is shown in Fig.~\ref{XTEJ1752}.  The fluxes for XTE J1752-223 in Table~\ref{Flux_table} are integrated over the days when XTE J1752-223 was observed to be in a high gamma-ray intensity state, MJD 55129--55218.  

RXTE measurements indicate a black hole low/hard spectrum with a power law component ($\Gamma \sim 1.4$) superimposed on a weak black body (kT $\sim 0.8$ keV). A 6.4 keV iron line is also seen, with combined spectral and timing properties similar to those observed in Cyg X-1 and GX 339-4 \citep{Shaposhnikov2009}. Results from {\it RXTE}/HEXTE analysis have shown evidence for emission up to 200 keV \citep{Munoz-Darias2010b}, best fit with a broken power law with a break energy near 130 keV, again markedly similar to Cyg X-1.  The flux ratios measured with GBM ($R_{50} = 0.218 \pm 0.005, R_{100} = 0.090 \pm 0.005, R_{300} = 0.006 \pm 0.006$) are similar to those observed for Cyg X-1 and are consistent with a $\Gamma \sim 1.7$ spectrum steepening to at least $\Gamma \sim 1.9$ above 100 keV.  

\subsubsection{GX 339-4}

\begin{figure}[t]
\includegraphics[width=78mm]{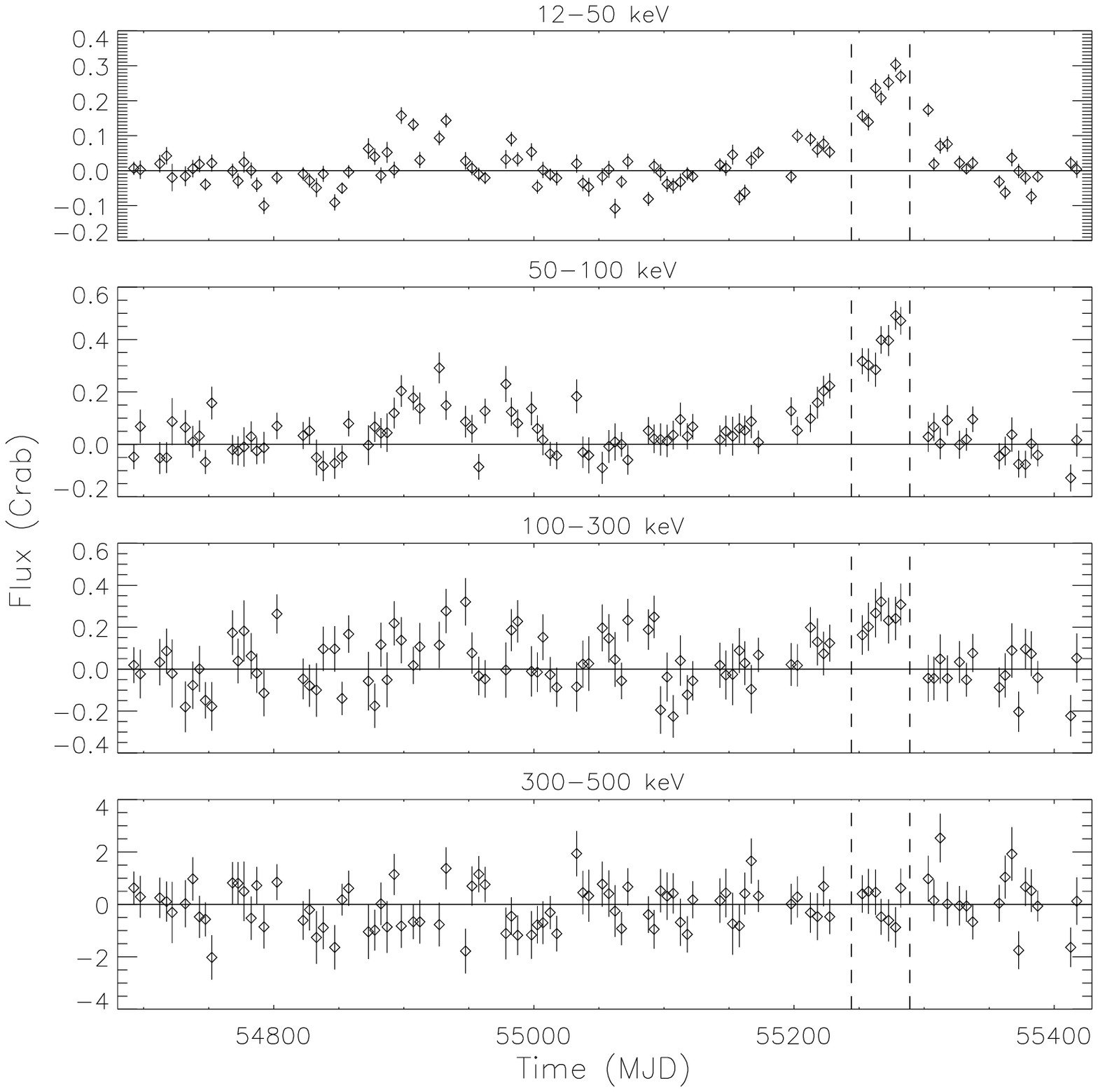}%
\caption{\label{GX339} The GBM light curve for GX 339-4. The light curve has been binned 5 days per data point.  The vertical dashed lines at MJD 55244 and MJD 55289 mark the flaring region used to derive the average fluxes in Table~\ref{Flux_table}. The fluxes are in Crab units, and the horizontial solid lines mark the zero flux levels.}
\end{figure}

The highly variable LMXB and black hole candidate GX 339-4 \citep{Samimi1979,Doxsey1979} is characterized by rapid time variability and low/hard X-ray states similar to those of Cyg X-1 \citep{Harmon1994,Cowley2002}. The results of analysis of both BATSE \citep{Case2008} and {\it INTEGRAL} \citep{Garcia2009} data have indicated the presence of high energy emission above 200 keV during previous outbursts, with the {\it INTEGRAL} spectrum fitted by a thermal Comptonization component together with synchrotron or self-synchrotron emission possibly originating
at the base of a jet. 

GX 339-4 was observed by MAXI to begin a large flare event starting on 2010 January 3 \citep{Yamaoka2010}. The flux observed by GBM began to increase starting in early 2010 January and continued to increase up to a level of $\sim400$ mCrab (12--25 keV), $\sim650$ mCrab (25--50 keV), $\sim 800$ mCrab (50--100 keV), and $\sim 550$ mCrab (100--300 keV) by early-April 2010, after which it began to rapidly decrease. It returned to quiescence in the higher energy bands by mid-April and in the 12--50 keV band by the end of April. The fluxes for GX 339-4 in Table~\ref{Flux_table} are integrated over the days when GX 339-4 was observed to be in a high gamma-ray state, MJD 55244--55289. Similar to Cyg X-1 and XTE J1752-223, the flux ratios measured by GBM ($R_{50} = 0.223  \pm 0.012$ and $R_{100} = 0.075 \pm 0.011$, with no measurable intensity above 300 keV) appear consistent with a $\Gamma \sim 1.7$ power law steepening above 100 keV to at least $\Gamma \sim 1.9$.

Note that there was a weaker double-peaked flare starting around MJD 54888 \citep{Markwardt2009a} and lasting until around MJD 55000 (see Fig.~\ref{GX339}). While the light curve in the 100--300 keV band is suggestive of positive emission in this band, it is only marginally significant ($\sim5\sigma$) with an average flux of $162 \pm 31$ mCrab.

\section{Conclusions and Future Prospects \label{sec:dis}}

Using the Earth occultation technique, the GBM instrument on \fermi has been monitoring the gamma-ray sky in the $\sim8-1000$ keV energy range, providing daily measurements for a catalog of 82 sources, including 72 x-ray binaries, five AGN, two magnetars, two cataclysmic variables, the Crab, and the Sun.  After the first two years of the \fermi mission, the Earth occultation technique applied to the GBM CTIME data has been used to detect six persistent sources and two transients at energies above 100 keV with a significance greater than $7\sigma$, demonstrating the capability of GBM to observe and monitor such sources. Two of the sources, the Crab and Cyg X-1, were detected with high significance above 300 keV. 

Light curves of all eight sources were presented in four broad energy bands from 12--500 keV.  The outbursts from the transient sources XTE J1752-223 and GX 339-4 were clearly visible in the 12--50 keV, 50--100 keV, and 100--300 keV broad bands. XTE J1752-223 was a previously unknown source, and the GBM light curves in the hard x-ray/low energy gamma-ray energy bands are consistent with the initial classification of this object as a black hole candidate in a bright low/hard state. The steep decline of the hard x-ray emission starting around MJD 55215 corresponded to an increase in the soft x-ray flux \citep{Homan2010,Negoro2010a}, indicating the transition from the low/hard state to a soft state. When XTE J1752-223 returned to the low/hard state around MJD 55282 \citep{Munoz-Darias2010a}, the hard x-ray emission was below the sensitivity limit of GBM. 

The hard emission seen from GX 339-4 is consistent with the bright hard states seen in previous outbursts from this object.  

Monitoring of Cyg X-1 at the onset of a recent state transition showed a steady decrease in the 100--300 keV flux that began about 19 days before the soft x-ray flux began to rise. As of MJD 55419, Cyg X-1 remains in a soft state, and we will continued to monitor Cyg X-1 in anticipation of the transition back to the canonical hard state.

While the GBM CTIME data used here does not have enough spectral resolution to produce detailed spectra, the flux ratios between the 12--50 keV broad band and the 50--100, 100--300, and 300--500 keV broad bands for the Crab are generally consistent with those inferred from the measured {\it INTEGRAL} spectrum, with the GBM results suggesting a slightly harder spectrum. The flux ratios observed for the transient sources XTE J1752-223 and GX 339-4 are similar to Cyg X-1 when it is in its canonical low/hard state, again consistent with these transients being observed in bright low/hard states. Future work will use the GBM CSPEC data, with its finer energy binning, to examine the detailed spectra for all of these sources, with particular emphasis on the low energy gamma-ray energy range. Also, the BGO detectors, with their greater sensitivity at higher energies, will be used to obtain additional measurements at energies above 150 keV. Several of the detected sources have spectral breaks or cutoffs in the 100--300 keV range, and we will look for these features and monitor their evolution over time.

We will continue to add to the list of sources being monitored as appropriate. We have detected Cen A with GBM in all energy bands up to 300 keV. BATSE detected several BL Lac objects known to exhibit flaring activity on the time scale of days \citep{Connaughton1999}.  The flaring behavior of blazars is sometimes accompanied by a shift upwards in the peak of the synchroton spectrum (e.g. above 100 keV in the 1997 Mrk 501 flare \citep{Petry2000}), making detection of these sources possible with GBM. While none have been detected so far, we will continue to expand our monitoring program to include the blazars with flares detected in other wavelengths.  We plan to expand our monitoring program to include other AGN as well.

We continue to fine tune the algorithms and work to reduce the systematic errors in the flux determination.  In the case of the BATSE Earth occultation analysis, there was clear evidence for the presence of sources in the data which were not in the occultation input catalog and which caused uncertainty in the assignment of fluxes to the sources that were in the input catalog. We see similar evidence for these uncataloged sources with GBM, especially below about 50 keV. To address this, our approach is two-fold: (1) We compare our GBM measurements with overlapping energy bands from other operating missions and regularly update our catalog as new transient outbursts are observed with GBM or other instruments. Light curves are regenerated if needed after catalog updates. (2) We are developing an imaging technique for GBM to produce an all-sky map of hard X-ray/soft gamma-ray sources. This map will then be used to identify sources not currently in the GBM occultation catalog, to expand the catalog, and to reduce the uncertainties in the measured fluxes. 


\acknowledgments

This work is supported by the NASA Fermi Guest Investigator program.  At LSU, additional support is provided by NASA/Louisiana Board of Regents Cooperative Agreement NNX07AT62A. A.C.A. wishes to thank the Spanish Ministerio de Ciencia e Innovaci\'on for support through the 2008 postdoctoral program MICINN/Fulbright under grant 2008-0116.

\end{document}